\documentstyle[sprocl,epsfig]{article}
\newcommand{\MeV}{{\rm\,MeV}}
\newcommand{\GeV}{{\rm\,GeV}}
\newcommand{\slk}{/\kern-6pt k}
\newcommand{\slp}{p\kern-5pt/}

\newcommand{\pfrac}[2]{\left(\frac{#1}{#2}\right)}
\title{TOP QUARK PRODUCTION NEAR THRESHOLD AT NLC
\footnote{Talk given by O.Yakovlev at the Conference 
``PHENOMENOLOGY-2000'', \\ Madison, Wisconsin, April 15-17, 2000}}
\author{OLEG YAKOVLEV}
\address{Randall Laboratory of Physics, University of Michigan,\\ 
Ann Arbor, Michigan 48109-1120, USA}
\author{STEFAN GROOTE}
\address{Institut f\"ur Physik der Johannes-Gutenberg-Universit\"at,\\
  Staudinger Weg 7, D-55099 Mainz, Germany\\[12pt]
 Floyd R. Newman Laboratory, Cornell University,\\
  Ithaca, NY 14853, USA}
\begin{document}

\maketitle\abstracts{In this talk we discuss the process $e^+e^-\to t\bar t$ 
near threshold. In particular we discuss a quark mass definition which is a
generalization of the static PS mass. The new definition allows us to
calculate recoil corrections to the static PS mass. Using this result we
calculate the cross section of $e^+e^-\to t\bar t$ near threshold at NNLO
accuracy adopting three alternative approaches, namely (1) fixing the pole
mass, (2) fixing the PS mass, and (3) fixing the new mass which we call the
$\overline{\rm PS}$ mass. We demonstrate that perturbative predictions for the
cross section become much more stable if we use the PS or the
$\overline{\rm PS}$ mass for the calculations. A careful analysis suggests
that the top quark mass can be extracted from a threshold scan at NLC with an
accuracy of about $100-200\MeV$.}
\centerline{\footnotesize (Preprint: UM-TH-00-22, CLNS 00/1690)}
\section{Introduction} 
\subsection{ On NLC, FMC, LHC and Tevatron}
The top quark physics will be one of the main subject of future $e^+e^-$ and
$\mu^+\mu^-$ colliders such as the Next Linear Collider (NLC) and the Future
Muon Collider (FMC). The goals are to measure and to determine the properties 
of the top quark which was first discovered at the Tevatron~\cite{Fermi} with
a mass of $m=174.3\pm 5\GeV$~\cite{Fermi1}. Although the top quark will be
studied at the LHC and the Tevatron (RUN-II) with an expected accuracy for the
mass of $2-3\GeV$, the {\em most accurate measurement\/} of the mass with an
accuracy of $0.1\%$ ($100-200\MeV$) is expected to be obtained only at the 
NLC~\cite{Peskin}. 

\subsection{Why do we need $10^{-3}$ accuracy in the top mass?}
The top quark mass enters the relation between the electroweak precision
observables indirectly through loops effects. The global electro-weak fit
of the Standard Model requires to have very accurate input data in order to
make a constraint for the masses of undiscovered particles, such as the Higgs
boson or other particles. The increase in the accuracy of the top quark mass
will improve the limits on the Higgs mass. In addition we can study possible
deviations from the Standard Model through anomalous couplings, CP violation,
or extra dimensions.

\subsection{LO and NLO cross section}
Due to the large top quark width, the top-antitop pair cannot hadronize into
toponium resonances. The cross section appears therefore to have a smooth
line-shape showing only a moderate $1S$ peak. In addition the top quark width
serves as an infrared cutoff~\cite{Fadin:1987wz,Fadin:1988fn} and as a natural
smearing over the energy~\cite{Poggio:1976af}. As a result, the nonperturbative
QCD effects induced by the gluon condensate are
small~\cite{Strassler:1991nw,Fadin:1991jh}, allowing us to calculate the cross
section with high accuracy by using perturbative QCD even in the threshold
region. Many theoretical studies at LO and NLO have been done in the past for
the total cross
section~\cite{Fadin:1987wz,Fadin:1988fn,Strassler:1991nw,Kwong:1991iy}, 
for the momentum distribution~\cite{Jezabek:1992np,Sumino:1993ai}, also
accounting for electro-weak corrections~\cite{Guth,Hollik,FadYak1}, and for
the complete NLO correction including non-factorizable
corrections~\cite{Melnikov:1994np,Fadin:1994dz,Fadin:1994kt,SuminoTH}.
It was proven that the non-factorizable corrections 
cancel in the inclusive cross section at 
NLO~\cite{Melnikov:1994np,Fadin:1994dz,Fadin:1994kt,Melnikov:1995hr,Melnikov:1994av,Melnikov:1996fx}.  

\subsection{On NNLO results}
Recently, the NNLO analysis of the inclusive threshold production cross
section has been reported~\cite{Hoang:1998xf,Melnikov:1998pr,Yakovlev:1999ke,Beneke:1999qg,Hoang:1999zc,Nagano,Penin}. The results of the
NNLO analysis are summarized in a review article~\cite{Review}. To summarize
the results for a standard approach using the pole mass, the NNLO corrections
are uncomfortably large, spoiling the possibility for the top quark mass
extraction at NLC with good accuracy because the $1S$ peak is shifted by about
$0.5\GeV$ by the NNLO, the last known correction. One of the main reasons for
this is the usage of the pole mass in the calculations. It was realized that
such type of instability is caused by the fact that the pole mass is a badly
defined object within full QCD~\cite{Beneke:1994sw,Vainshtein}.

In this talk we discuss a definition of the quark mass alternative to the pole
mass. It is the so-called potential subtracted (PS) mass, suggested in
Ref.~\cite{Beneke}. In contrast to the pole mass the PS mass is not sensitive
to the non-perturbative QCD effects. We derive recoil corrections to the
relation of the pole mass to the PS mass and demonstrate that perturbative
predictions for the cross section become much more stable at higher orders of
QCD (shifts are below $0.1\GeV$) if we use the PS mass for the calculations. 
This understanding removes one of the obstacles for the accurate top quark
mass measurement and it can be expected that the top quark mass will be
extracted from a threshold scan at NLC with an accuracy of about $100-200\MeV$.
We have to mention that the necessity of isolating the IR contributions in
the mass calculation and the consideration of a short distance mass have been
studied intensively~\cite{Vainshtein,Beneke,Uraltsev1,Uraltsev2,Hoang1}. The 
applications of the PS, LS and $1S$ mass have been reported recently by
several groups~\cite{Beneke:1999qg,Hoang:1999zc,Nagano,Karlsruhe,MelnikovK}
and reviewed in Ref.~\cite{Review}. The detailed comparison of our results
using the PS mass with results of other groups have been performed in
Ref.~\cite{Review} so that we refer the reader to this reference for details.

In this talk we review Ref.~\cite{YG}. In particular, we discuss a quark mass
definition (we call it $\overline{\rm PS}$ mass) which is a generalization of
the static PS mass proposed by M.~Beneke~\cite{Beneke}. The new definition
allows us to calculate recoil corrections to the static PS mass. Second, we
calculate the cross section for $e^+e^-\to t\bar t$ near threshold with NNLO
accuracy using three alternative mass schemes, i.e.\ (1) the pole mass, (2) 
the static PS mass, and (3) the new $\overline {\rm PS}$ mass. The results of
the first approach has already been reported in Ref.~\cite{Yakovlev:1999ke},
the results for the top quark pair production near threshold using the static
PS mass were presented in Refs.~\cite{Karlsruhe,MelnikovK,YG,Review}, the third
alternative schemes were discussed in Refs.~\cite{YG}.

\section{Top quark production near threshold\\ with NNLO accuracy}
In this section we consider the cross section of the process
$e^+e^-\to t\bar t$ in the near threshold region where the velocity $v$ of
the top quark is small. It is well known that the conventional perturbative 
expansion does not work in the non-relativistic region because of the presence
of the Coulomb singularities at small velocities $v\to 0$. The terms
proportional to $(\alpha_s/v)^n$ appear due to the instantaneous Coulomb
interaction between the top and the antitop quark. The standard technique for
re-summing the terms $(\alpha_s/v)^n$ consists in using the Schr\"odinger
equation for the Coulomb potential and in finding the Green 
function~\cite{Fadin:1987wz,Fadin:1988fn}. The Green function is then related
to the cross section by the optical theorem. In order to determine NLO
corrections to the cross section we need to know the short-distance correction
to the vector current~\cite{NLOcurrent}, the NLO correction to the Coulomb
potential~\cite{Billoire:1980ih}, and the contribution of the non-factorizable
corrections~\cite{Melnikov:1994np,Fadin:1994dz,Fadin:1994kt}. It was proven
that the non-factorizable corrections cancel in the inclusive cross section at
NLO and beyond~\cite{Melnikov:1994np,Fadin:1994dz,Fadin:1994kt,Melnikov:1995hr,Melnikov:1994av,Melnikov:1996fx}. 
At NNLO the situation is more complicated. One of the obstacles for a
straightforward calculation are the UV divergences coming from relativistic
corrections to the Coulomb potential. This problem can be solved by a proper
factorization of the amplitudes and by employing effective theories. We want
to sketch the derivation of the inclusive cross section. The inclusive cross
section can be obtained from the correlation function of two vector currents
$j_{\mu}(x)=\bar t(x)\gamma_\mu t(x)$,
\begin{equation}
\Pi_{\mu\nu}(p^2)=i\int d^4xe^{ip\cdot x}
  \langle 0|T\{j_\mu(x),j_\nu(0)\}|0\rangle. 
\end{equation}
The first step is to expand the Lagrangian and the currents
\begin{equation}
j_i=\bar t\gamma_it=c_1\psi^\dagger\sigma_i\chi
  -\frac{c_2}{6m^2}\psi^\dagger\sigma_i(i{\bf D})^2\chi
\end{equation}
consistently in $1/m$. The useful language for treating the one and two
non-relativistic quark(s) is provided by the
NRQCD~\cite{Caswell:1986ui,Bodwin:1995jh} and the PNRQCD~\cite{Pineda},
respectively. After the expansion, the cross section reads
\begin{eqnarray}\label{cross}
R&=&\sigma(e^+e^-\to t\bar t)/\sigma_{pt}\nonumber\\
  &=&e^2_QN_c\frac{24\pi}sC(r_0){\rm Im}\left[\left(1-\frac{\vec p\,^2}{3m^2}
  \right)G(r_0,r_0|E+i\Gamma)\right]\Bigg|_{r_0\to 0}
\end{eqnarray}
where $\sigma_{pt}=4\pi\alpha^2/3s$, $e_Q$ is the electric charge of the top
quark, $N_c$ is the number of colors, $\sqrt{s}=2m+E$ is the total
center-of-mass energy of the quark-antiquark system, $m$ is the top quark pole
mass and $\Gamma$ is the top quark width. The unknown coefficient $C(r_0)$ can
be fixed by using a direct QCD calculation of the vector vertex at NNLO in the
so-called intermediate region~\cite{Czarnecki:1998vz,Beneke:1998zp} and by
using the direct matching procedure suggested in Ref.~\cite{HoangM}.

The function $G(\vec r,\vec r\,'|E+i\Gamma)$ is the non-relativistic Green
function. It satisfies the Schr\"odin\-ger equation
\begin{equation}
(H-E-i\Gamma)G(\vec r,\vec r\,'|E+i\Gamma)=\delta(\vec r-\vec r\,'),
\end{equation}
$H$ is the non-relativistic Hamiltonian of the heavy quark-antiquark system.
It is shown~\cite{Hoang:1998xf,Melnikov:1998pr,Yakovlev:1999ke} that the
Schr\"odinger equation can be reduced to the equation
\begin{equation}\label{se1}
(H_1-E_1)G_1(r,r'|E_1)=\delta^3(r-r')
\end{equation}
with the energy $E_1=\bar E+\bar E^2/4m$, $\bar E=E+i\Gamma$, and with the
Hamiltonian 
\begin{eqnarray}
\lefteqn{H_1\ =\ \frac{\vec p\,^2}{m}+V_C(r)+\frac{3\bar E}{2m}V_0(r)
  -\left(\frac23+\frac{C_A}{C_F}\right)\frac{V_0^2(r)}{2m},}\nonumber\\
\lefteqn{V_0(r)\ =\ -\frac{\alpha_s(\mu)C_F}r,}\nonumber\\
\lefteqn{V_C(r)\ =\ V_0(r)\Bigg\{1+\frac{\alpha_s(\mu)C_F}{4\pi}
  \left(2\beta_0\ln(\mu'r)+a_1\right)}\\&&
  +\pfrac{\alpha_s(\mu)}{4\pi}^2\left(\beta_0^2
  \left(4\ln^2(\mu'r)+\frac{\pi^2}3\right)+2(\beta_1+2\beta_0a_1)\ln(\mu'r)
  +a_2\right)\Bigg\}\nonumber
\end{eqnarray}
where $\mu'=\mu e^{\gamma_E}$, $\mu$ is the renormalization scale, and
$\gamma_E$ is Euler's constant. The QCD beta function coefficients $\beta_0$
and $\beta_1$ and the coefficients $a_1$ and $a_2$ are listed later this talk.
The final expression for the NNLO cross section is given by
\begin{eqnarray}\label{final}
R^{\rm NNLO}(E)&=&\frac{8\pi}{m^2}\left\{1-4C_F\frac{\alpha_s(m)}\pi
  +C_2(r_0)C_F\pfrac{\alpha_s(m)}\pi^2\right\}\
  \times\nonumber\\&&\qquad\times\
  {\rm Im}\left[\left(1-\frac{5\bar E}{6m}\right)G_1(r_0,r_0|E_1)\right]
\end{eqnarray}
with $G_1(r_0,r_0|E_1)$ being the solution of Eq.~(\ref{se1}) at $r=r'=r_0$.
For the numerical solution we use the program derived in
Ref.~\cite{Yakovlev:1999ke} by one of the authors.

\section{On the mass definitions}
The top quark mass is an input parameter of the Standard Model. Although it is
widely accepted that the quark masses are generated due to the Higgs mechanism,
the value of the mass cannot be calculated from the Standard Model. Instead,
quark masses have to be determined from the comparison of theoretical
predictions and experimental data. 

It is important to stress that there is no unique definition of the quark mass.
Because the quark cannot be observed as a free particle like the electron,
the quark mass is a purely theoretical notion and depends on the concept
adopted for its definition. The best known definitions are the pole mass and
the $\overline {\rm MS}$ mass. However, both definitions are not adequate for 
the analysis of top quark production near threshold. The pole mass should not
be used because it has the renormalon ambiguity and cannot be determined more
accurately than $300-400\MeV$~\cite{Beneke:1994sw,Vainshtein} (see also
Refs.~\cite{Beneke,Scot,SuminoRen}). Actually, we may relate the pole mass
with some short distance mass like the $\overline{\rm MS}$ mass $M$ which for
the ``large $\beta_0$ approximation'' (see e.g.\ Ref.~\cite{AgliettiLigeti})
is given by
\begin{equation}
m_{\rm pole}
  =M\left\{1+\sum m_n\left(\frac{\beta_0\alpha_s}{4\pi}\right)^n\right\}.
\end{equation}
The Borel image of the pole mass $m_{\rm pole}$ reads  
\begin{equation}
\bar m_{\rm pole}(u)=\sum m_n\frac{u^n}{n!}
\end{equation}
The function $m_{\rm pole}(u)\sim\Gamma(1-2u)$ has poles at real and positive
values $u=1/2,3/2,\ldots$. Therefore, the inverse Borel transformation
\begin{equation}
m_{\rm pole}=\int du\exp\left(\frac{4\pi u}{\beta_0\alpha_s}\right)
  \bar m_{\rm pole}(u)
\end{equation}
generates ambiguities of the order
\begin{equation}
\delta m\approx\exp\left(\frac{4\pi}{\beta_0\alpha_s}\right)
  \approx\Lambda_{\rm QCD}.
\end{equation}
These ambiguities are known in the literature as renormalon ambiguities (see
for example Refs.~\cite{BenekeRen,PivovarovRen}). 

The $\overline{\rm MS}$ mass is an Euclidean mass, defined at high virtuality,
and therefore destroys the non-relativistic expansion. Instead, it was
recently suggested to use threshold masses like the low scale (LS)
mass~\cite{Vainshtein}, the potential subtracted (PS) mass~\cite{Beneke}, or
one half of the perturbative mass of a fictious $1^3S_1$ ground state (called
$1S$ mass)~\cite{Hoang:1999zc}. In this talk we focus on the PS mass suggested
in Ref.~\cite{Beneke},
\begin{eqnarray}\label{defPS}
m_{\rm PS}=m_{\rm pole}-\delta m_{\rm PS}\quad\mbox{with}\quad
\delta m_{\rm PS}=-\frac12\int^{|\vec k|<\mu_f}\frac{d^3k}{(2\pi)^3}
  V_C(|\vec k|)
\end{eqnarray}
where $V_C$ is the quark-antiquark Coulomb potential.

\subsection{On renormalons and soft QCD effects cancellation}
In order to understand why this mass is better defined than the pole mass and
to see that the pole mass is very sensitive to long distance effects, it is
enough to consider the one-loop expression for the self energy diagram. Taking
the residue in $k_0$, one obtains a soft self energy contribution which comes
from momenta $k$ with $|\vec k|<\Lambda_{\rm QCD}$,
\begin{equation}\label{critical}
\delta m=4\pi\alpha_sC_F\kern-12pt
  \int\limits_{|\vec k|<\Lambda_{\rm QCD}}\kern-12pt
  \frac{d^3k}{(2\pi)^3}\frac1{2|\vec k|^2}
  =\frac{\alpha_s C_F}{\pi}\Lambda_{\rm QCD}.
\end{equation}
We observe that the pole mass has a non-perturbative uncertainty of order
$\Lambda_{\rm QCD}$ which then penetrates into consequent perturbative 
QCD calculations. This uncertainty or ambiguity is caused by the poles of the
Borel transformation as shown before. However, it is easy to realize that the
PS mass is free of this ambiguity. Indeed, the term $\delta m$ in
Eq.~(\ref{critical}) and therefore the renormalon ambiguities associated with
it cancel in the definition of the PS mass as given in Eq.~(\ref{defPS}) as
well as in the combination $2m_{\rm pole}+V(r)$ which is known as static
energy. We can see this by applying the Borel technique also to the QCD
potential 
\begin{equation}
V(r)=\int\frac{d^3 k}{(2\pi)^3}\left(-\frac{4\pi C_F\alpha_s(k^2)}{k^2}\right)
  e^{ik\cdot r}.
\end{equation}
The Borel image of the potential
$V(r)=\sum V_n(\beta_0\alpha_s/4\pi)^n$ in the ``large $\beta_0$ limit'' is
given by
\begin{equation}
\bar V(u)=\sum V_n\frac{u^n}{n!}.
\end{equation}
with the function $\bar V(u)\sim\Gamma(1-2u)$ having again poles at real and
positive values $u=1/2,3/2,\ldots$. The inverse Borel transformation
\begin{equation}
V(r) =\int du\exp\left(\frac{4\pi u}{\beta_0\alpha_s}\right)\bar V(u).
\end{equation}
generates ambiguities of the order
\begin{equation}
\delta V(r)\approx\exp\left(\frac{4\pi}{\beta_0\alpha_s}\right)
  \approx\Lambda_{\rm QCD}.
\end{equation}
The coefficient in $\bar V(u)$ differs from $\bar m_{\rm pole}(u)$ by a factor
$-2$. Therefore we observe the cancellation of the renormalon ambiguity in  
the combination $2m_{\rm pole}+V(r)$~\cite{Beneke,BenekeRen,Scot}.

The definition in Eq.~(\ref{defPS}) has been given in Ref.~\cite{Beneke} but
has already been discussed implicitly in Ref.~\cite{Vainshtein}. The
remarkable step made in Ref.~\cite{Beneke} is to use this definition beyond
one-loop order. It has been proven in Ref.~\cite{Beneke} that the cancellation
of the infrared QCD contributions to the PS mass in Eq.~(\ref{defPS}) holds
even at higher loop orders. Still the definition in Eq.~(\ref{defPS}) is valid
only in the static approximation and does not contain $1/m$ corrections like
the other threshold mass definitions. We therefore suggest to extend the
definition given in Ref.~\cite{Beneke}. 

\subsection{The $\overline{\rm PS}$ mass definition}
Our objective is that the definition should be gauge independent and
well-defined within quantum field theory so that radiative and relativistic
corrections can be calculated in a systematic way. Our definition is given by
\begin{eqnarray}
m_{\overline{\rm PS}}=m_{\rm pole}-\delta m_{\overline{\rm PS}}
  \quad\mbox{with}\quad
  \delta m_{\overline{\rm PS}}=\Sigma_{\rm soft}(\slp)\Big|_{\slp=m}, 
\end{eqnarray}
where $\Sigma_{\rm soft}$ is the soft part of the heavy quark self energy
which is defined as the part where at least one of the heavy quark propagators
is on-shell.
\begin{figure}
\begin{center}
\epsfig{figure=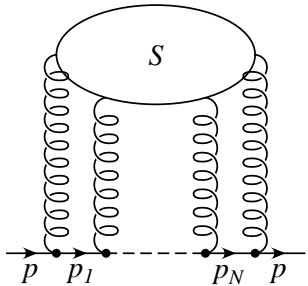, scale=0.4}
\caption{\label{fig1}
  The general structure of the self energy diagram of a quark} 
\end{center}
\end{figure}
Starting point for our considerations is therefore the self energy of an
on-shell quark with mass $m$ and momentum $p$ (i.e.\ $p^2=m^2$) as shown in
Fig.~\ref{fig1} and written down as
\begin{eqnarray}\label{def1}
-i\Sigma(\slp)&=&\int\prod_{m=1}^M\frac{d^4l_m}{(2\pi)^4}
  S^{\{a_n\}}_{\{\alpha_n\}}(\{l_m\})\left(-ig_s\gamma^{\alpha_{N+1}}
  T_{a_{N+1}}\right)\ \times\nonumber\\&&\qquad\qquad\qquad\times\
  \prod_{n=N}^1\frac{i}{\slp_n-m}\left(-ig_s\gamma^{\alpha_n}T_{a_n}\right)
\end{eqnarray}
where the last factor is a non-commutative product with decreasing index $n$.
The line momenta $k_n$ are linear combinations of the gluon loop momenta $l_m$,
the particular representation is specified by the structure $S$. The symbol
$\{l_m\}$ means the set of all these loop momenta, the same symbol is used
for the Lorentz and color indices. In general we have $M\le N$ which means
that line momenta can be correlated. The momenta of the virtual quark states
are given by $p_n=p+k_n$.

The quark is considered to be at rest, $p=(m,\vec 0)$, and it interacts with a
number of gluons. The subdiagram $S$ displayed in Fig.~\ref{fig1} describes
the interaction between the gluons. In general the quark lines between the
interaction points represent virtual quark states. However, if the virtual
quark comes very close to the mass shell and the total momentum of the cloud
of virtual gluons becomes soft, this situation gives rise to long-distance
nonperturbative QCD interactions. The described (virtual) contributions
results in the soft part of the self energy, $\Sigma_{\rm soft}$. So we define
\begin{eqnarray}\label{def2}
\lefteqn{-i\Sigma_{\rm soft}(\slp)\ =\ \sum_{i=1}^N\int\prod_{m=1}^M
  \frac{d^4l_m}{(2\pi)^4}S^{\{a_n\}}_{\{\alpha_n\}}(\{l_m\})
  \left(-ig_s\gamma^{\alpha_{N+1}}T_{a_{N+1}}\right)\
  \times}\nonumber\\&&\times\ \prod_{n=N}^{i+1}
  \frac{i}{\slp_n-m}\left(-ig_s\gamma^{\alpha_n}T_{a_n}\right)
  i(\slp_i+m)\left(-i\pi\delta(p_i^2-m^2)\right)\
  \times\nonumber\\&&\qquad\times\
  \left(-ig_s\gamma^{\alpha_i}T_{a_i}\right)\prod_{n=i-1}^1
  \frac{i}{\slp_n-m}\left(-ig_s\gamma^{\alpha_n}T_{a_n}\right).
\end{eqnarray}
One can derive this expression from Eq.~(\ref{def1}) by using the identity
\begin{equation}\label{ident1}
\frac1{p^2-m^2+i\epsilon}=-i\pi\delta(p^2-m^2)+P\pfrac1{p^2-m^2}
\end{equation}
and the fact that the principal value integral does not give any infrared
sensitive contribution. The delta function can be used to remove the
integration over the zero component of $k_i$. In order to parameterize the
softness of the gluon cloud we impose a cutoff on the spatial component,
$|\vec k_i|<\mu_f$, and indicate this by a label $\mu_f$ written at the upper
limit of the three-dimensional integral. This cutoff $\mu_f$ is also known as
{\em factorization scale\/}. So we can rewrite Eq.~(\ref{def2}) as
\begin{equation}
\Sigma_{\rm soft}(\slp,\mu_f)=-\frac12\sum_{i=1}^N\int^{\mu_f}
  \frac{d^3k_i}{(2\pi)^3}V(\vec k_i,p)
\end{equation} 
where
\begin{eqnarray}
\lefteqn{V(\vec k_i,p)\ =\ -\int\prod_{m=1}^{M-1}\frac{d^4l_m}{(2\pi)^4}
  S^{\{a_n\}}_{\{\alpha_n\}}(\{l_m\})\ \times}\nonumber\\&&\times\ 
  \left(-ig_s\gamma^{\alpha_{N+1}}T_{a_{N+1}}\right)\prod_{n=N}^{i+1}
  \frac{i}{\slp_n-m}\left(-ig_s\gamma^{\alpha_n}T_{a_n}\right)
  \frac{\slp_i+m}{2p_i^0}\ \times\nonumber\\&&\times\
  \left(-ig_s\gamma^{\alpha_i}T_{a_i}\right)\prod_{n=i-1}^1
  \frac{i}{\slp_n-m}\left(-ig_s\gamma^{\alpha_n}T_{a_n}\right).
\end{eqnarray}
The range of the index $m$ is reduced by one which indicates that one of the
loop momenta is extracted as line momentum of the $i$-th line. In the following
we deal with the different realizations of this compact expression. As we will
see explicitly, the function $V(\vec k,p)$ occurring as integrand can be seen
as quark-antiquark potential where we have summed over the spin of the tensor
product of a final state with an initial state. Because the static
quark-antiquark potential is used in a similar way in Ref.~\cite{Beneke}, we
recover the result of Ref.~\cite{Beneke} in the static limit.

\subsection{The one-loop contribution}
\begin{figure}
\begin{center}
\epsfig{figure=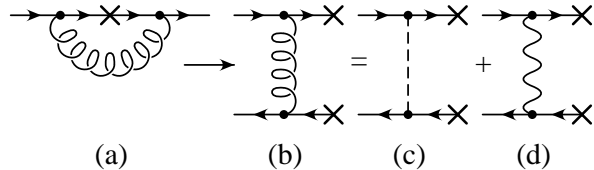, scale=0.4}
\caption{\label{fig2}Leading order contribution to the quark self energy
(a) and to the quark-antiquark potential (b). The cross indicates the
point where we cut the quark line by imposing an on-shell condition to
the virtual quark state. The gluon propagator can be decomposed in a
Coulomb propagator (c) and a transverse propagator (d).}
\end{center}
\end{figure}

The leading order contribution to the self energy of the quark is given by
\begin{equation}
\Sigma(\slp)=i\int\frac{d^4k}{(2\pi)^4}(-ig_s\gamma_\alpha T_a)
  \frac{i}{\slp+\slk-m}(-ig_s\gamma^\alpha T_a)\frac{-i}{k^2}
\end{equation}
where Feynman gauge is used for the gluon. The soft part of it is given by
\begin{equation}\label{leadsoft}
\Sigma_{\rm soft}(\slp,\mu_f)=-\frac12\int^{\mu_f}\frac{d^3k}{(2\pi)^3}
  V(\vec k,p),\quad V(\vec k,p)=V_+(\vec k,p)+V_-(\vec k,p)
\end{equation}
where
\begin{equation}\label{Vpm}
V_\pm(\vec k,p)=-\frac{g_s^2C_F(\sqrt{m^2+\kappa^2}\mp 2m)}{2m
  \sqrt{m^2+\kappa^2}(\sqrt{m^2+\kappa^2}-m)}
\end{equation}
and $\kappa=|\vec k|$. The procedure of taking the soft part by setting a cut
is shown in Fig.~\ref{fig2}(a--b).
The two potential parts are known as the scattering potential $V_+(\vec k,p)$
and the annihilation potential $V_-(\vec k,p)$ and correspond to the two
zeros $k_\pm=\pm\sqrt{\kappa^2+m^2}-m$ of the delta function. Integrating the
potential up to the factorization scale $\mu_f$, we obtain
\begin{equation}
\Sigma_{\rm soft}(\mu_f)=\frac{\alpha_sC_F}{2\pi} m
  \left\{3\ln\left(\frac{\mu_f}m+\sqrt{\frac{\mu_f^2}{m^2}+1}\,\right)
  -\frac{\mu_f}m\sqrt{\frac{\mu_f^2}{m^2}+1}\,\right\}.
\end{equation}
The expansion of this expression in small values of $\mu_f/m$ results in
\begin{equation}
\Sigma_{\rm soft}(\mu_f)=\frac{\alpha_sC_F}{\pi}\mu_f
  \left\{1-\frac{\mu_f^2}{2m^2}\right\}.
\end{equation}
The first term reproduces the result given in Ref.~\cite{Beneke} to leading
order in $\alpha_s$ while the second term is the recoil correction to the
static limit in this order of perturbation theory. This second term is related
to the Breit-Fermi potential but does not coincide with it. 

\subsection{Two loop contributions}
\begin{figure}
\begin{center}
\epsfig{figure=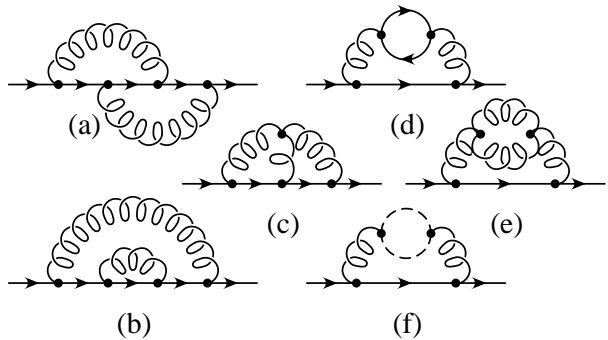, scale=0.4}
\caption{\label{fig3}Two-loop contributions to the quark self energy}
\end{center}
\end{figure}

To take a step beyond the leading order perturbation theory, we consider
two-loop diagrams for the heavy quark self energy as shown in Fig.~\ref{fig3}.
We calculate them in Coulomb gauge, even though we stress that our final
result is gauge invariant. The gluon propagator in Coulomb gauge is given by
\begin{equation}
G^{ab}_{00}(k)=\frac{i\delta^{ab}}{\vec k\,^2},\qquad
G^{ab}_{ij}(k)=\frac{i\delta^{ab}}{k^2}
  \left(\delta_{ij}-\frac{k_ik_j}{\vec k\,^2}\right),\qquad i,j=1,2,3.
\end{equation}
The use of Coulomb gauge splits up the gluon propagators into a Coulomb term
(Coulomb gluon) and a transverse term (transverse gluon) where the first one
couples to the quark via the time components only. This splitting is shown
in Fig.~\ref{fig2}(b--d).

In cutting the quark line in all possible ways we obtain a lot of diagrams
from the ones shown in Fig.~\ref{fig3}. However, we find that the final
contribution of the two abelian diagrams in Fig.~\ref{fig3}(a--b) to the
soft part of the self energy is suppressed by $\mu_f^2/m^2$. Therefore it
turns out that the only abelian diagrams which can give a non-suppressed
contribution to the soft part of the quark self energy are the diagrams
containing the vacuum polarization of the gluon as shown in
Fig.~\ref{fig3}(d--f). The simple calculation of these diagrams within the
$\overline{\rm MS}$ scheme, accounting only for light fermion loops, gluon
loop (and ghost loop if Feynman gauge is used) results after renormalization
in
\begin{eqnarray}
\Sigma_{\rm soft}^A&=&-\frac12\int\frac{d^3k}{(2\pi)^3}
  \left(\frac{4\pi C_F\alpha_s(\mu)}{|\vec k|^2}\right)\
  \times\nonumber\\&&\times\ \left\{1+\frac{\alpha_s}{4\pi}
  \left(\frac{31C_A}9-\frac{20T_FN_F}9-\left(\frac{11C_A}3-\frac{4T_FN_F}3
  \right)\ln\pfrac{|\vec k|^2}{\mu^2}\right)\right\}\nonumber\\
  &=&\frac{\alpha_S(\mu)C_F}\pi\mu_f\left\{1+\frac{\alpha_s(\mu)}{4\pi}
  \left(a_1-\beta_0\ln\pfrac{\mu_f^2}{\mu^2}\right)\right\}.
\end{eqnarray}
This result has been anticipated because the expression in the curly brackets
of the integrand reproduces the next-to-leading order correction to the QCD
Coulomb potential.
 
For the only non-abelian diagram in Fig.~\ref{fig3}(c) the use of Coulomb
gauge leads to seven diagrams. However, direct calculations show that only
the diagram where the transverse gluon is joined by Coulomb gluons on both
sides of the quark line contributes to the order $g_s^4\mu_f/m$. For this
diagram and therefore for the whole non-abelian contribution up to this
order we obtain
\begin{equation}
\Sigma_{\rm soft}^{NA}=\frac{\alpha^2_sC_FC_A}{8m}\mu_f^2.
\end{equation}
This result has been anticipated, too, to be minus one half of the non-abelian
correction to the QCD Coulomb potential, which is known in the literature 
(see for example Refs.~\cite{Gupta,Duncan}),
\begin{eqnarray}
\Sigma_{\rm soft}^{NA}&=&-\frac12\int^{\mu_f}\frac{d^3k}{(2\pi)^3}
  \left\{-\frac{\pi^2\alpha_s^2C_FC_A}{m|\vec k|}\right\}
  \ =\ \frac{\alpha_s^2C_FC_A}{8m}\mu_f^2.
\end{eqnarray}

\subsection{Our final result}
Summarizing all contribution up to NNLO accuracy, we obtain
\begin{eqnarray}\label{psmass}
m_{\overline{\rm PS}}(\mu_f)-m&=&-\frac{\alpha_s(\mu)C_F}\pi\mu_f
  \Bigg\{1+C'_0\frac{\mu_f}m+C''_0\frac{\mu_f^2}{m^2}\nonumber\\&&\qquad\qquad
  +\frac{\alpha_s(\mu)}{4\pi}\left(C_1+C'_1\frac{\mu_f}m\right)
  +C_2\pfrac{\alpha_s(\mu)}{4\pi}^2\Bigg\} 
\end{eqnarray}
where $m$ is the pole mass, $\mu$ is the renormalization scale, $\mu_f$ is
the factorization scale, and
\begin{eqnarray}
C_0&=&1,\qquad C'_0\ =\ 0,\qquad C''_0\ =\ -\frac12,\nonumber\\
C_1&=&a_1-2\beta_0\ln\pfrac{\mu_f}\mu,\qquad C'_1\ =\ C_A\frac{\pi^2}2,
  \nonumber\\
C_2&=&a_2-2(2a_1\beta_0+\beta_1)\left(\ln\pfrac{\mu_f}\mu-1\right)
  \nonumber\\&&\qquad
  +4\beta_0^2\left(\ln^2\pfrac{\mu_f}\mu-2\ln\pfrac{\mu_f}\mu+2\right).
\end{eqnarray}
The constants $a_1$ and $a_2$ are given by~\cite{Fischler,Peter,Schroder}
\begin{eqnarray}
a_1&=&\frac {31}9C_A-\frac{20}9T_FN_F,\nonumber\\
a_2&=&\left(\frac{4343}{162}+4\pi^2-\frac{\pi^4}4
  +\frac{22}3\zeta_3\right)C_A^2
  -\left(\frac{1798}{81}+\frac{56}3\zeta_3\right)C_AT_FN_F\nonumber\\&&
  +\left(\frac{20}9T_FN_F\right)^2
  -\left(\frac{55}3-16\zeta_3\right)C_FT_FN_F,
\end{eqnarray}
the coefficients of the beta function are known as
\begin{eqnarray}
\beta_0=\frac{11}3C_A-\frac43N_FT_F,\quad
\beta_1=\frac{34}3C_A^2-\frac{20}3C_AT_FN_F-4C_FT_FN_F
\end{eqnarray}
where $C_F=4/3$, $C_A=3$ and $T_F=1/2$ are color factors and $N_F=5$ is the
number of light flavors. The coefficients $C_1$ and $C_2$ in Eq.~(\ref{psmass})
have been derived in Ref.~\cite{Beneke} by using known corrections to the QCD
potential. In this work we have derived the coefficients $C'_0$, $C''_0$, and
$C'_1$. Note that our result can be represented in a condensed form as
\begin{equation}
m_{\overline{\rm PS}}(\mu_f)-m=-\frac12\int^{\mu_f}\frac{d^3k}{(2\pi)^3}
  \left(V_C(|\vec k|)+ V_R(|\vec k|)+V_{NA}(|\vec k|)\right)
\end{equation}
where the first term $V_C$ is the static Coulomb potential, $V_R$ is the
relativistic correction (which is related to Breit-Fermi potential but does
not coincide with it), and $V_{NA}$ is the non-abelian correction. Note that
we did not include electroweak corrections neither in this mass relation nor
in the cross section we will look on later. Finally (which is actually
{\em not\/} our result) the relation between the pole mass and the
$\overline{\rm MS}$ mass is given by the three-loop
relation~\cite{MelRit,Steinhauser}
\begin{eqnarray}\label{msmass}
\frac{m_{\rm pole}}{\overline{m}(\overline{m})}&=&1+\frac43\pfrac{\alpha_s}\pi
  +\pfrac{\alpha_s}\pi^2(-1.0414N_F+13.4434)\nonumber\\&&
  +\pfrac{\alpha_s}\pi^3(0.6527N_F^2-26.655N_F+190.595)
\end{eqnarray}
where $\alpha_s=\alpha_s(\overline{m})$ is taken at the $\overline{\rm MS}$
mass.

Using the relations between the masses, we fix the $\overline{\rm MS}$ mass to
take the values $\overline{m}(\overline{m})=160\GeV$, $165\GeV$, and $170\GeV$
and use Eq.~(\ref{msmass}) to determine the pole mass at LO, NLO, and NNLO.
This pole mass is then used as input parameter $m$ in Eq.~(\ref{psmass}) to
determine the PS and $\overline{\rm PS}$ masses at LO, NLO, and NNLO. The
obtained values for the PS and $\overline{\rm PS}$ mass differ only in NNLO.
The results of these calculations are collected in Table~\ref{tab1}, together
with the estimates for ``large $\beta_0$''
accuracy~\cite{AgliettiLigeti,BenekeBraun}. Note that the same values for
the $\overline{\rm MS}$ mass have been used for Tables~2 and~3 in
Ref.~\cite{Review}. The obtained mass values can now be used for the analysis
in the following section.

Taking the $\overline{\rm PS}$ mass instead of the PS mass, we observe an
improvement of the convergence. The differences for the mass values in going
from LO to NLO to NNLO to the ``large $\beta_0$'' estimate for e.g.\
$\overline{m}(\overline{m})=165\GeV$ read $7.64\GeV$, $1.64\GeV$, $0.52\GeV$,
and $0.25\GeV$ for the pole mass, $6.72\GeV$, $1.21\GeV$, $0.29\GeV$, and
$0.08\GeV$ for the PS mass and finally $6.72\GeV$, $1.21\GeV$, $0.27\GeV$, and
$0.08\GeV$ for the $\overline{\rm PS}$ mass.

\begin{figure}
\centerline{\epsfig{file=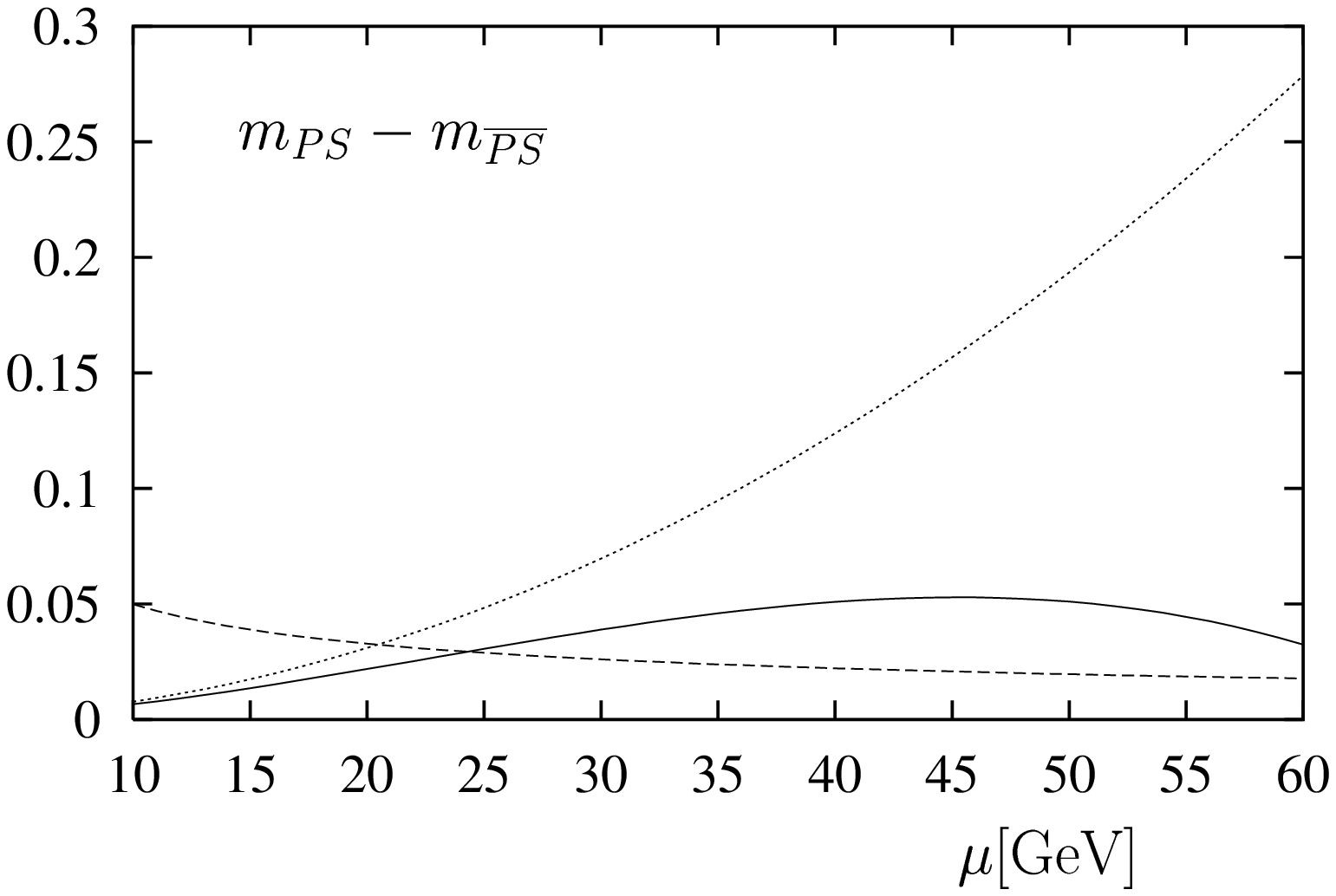,scale=0.7}}
\caption{\label{fig8}The difference between the PS and the $\overline{\rm PS}$
mass (in GeV) as a function of the factorization scale $\mu_f$ (solid line) at
$\mu=15\GeV$. The dotted line shows only the non-abelian part of the
difference. The dependence of $m_{\overline{\rm PS}}-m_{\rm PS}$ as a function
of the normalization scale $\mu$ at $\mu_f=30\GeV$ is shown as dashed line.}
\end{figure}

{\footnotesize 
\begin{table}[t]
\begin{center}
\begin{tabular}{|c||c|c|c|c||c|c|c|c|}\hline&&&&&&&&\\
$\overline{m}(\overline{m})$&$m_{\rm PS}^{\rm LO}$&$m_{\rm PS}^{\rm NLO}$&
$\displaystyle\matrix{m_{\rm PS}^{\rm NNLO}\cr
  m_{\overline{\rm PS}}^{\rm NNLO}}$&
$\displaystyle\matrix{m_{\rm PS}^{\beta_0}\cr
  m_{\overline{\rm PS}}^{\beta_0}}$&$m_{\rm pole}^{\rm LO}$&
  $m_{\rm pole}^{\rm NLO}$&
  $m_{\rm pole}^{\rm NNLO}$&$m_{\rm pole}^{\beta_0}$\\&&&&&&&&\\\hline
\hline &&&&&&&&\\
  160.0&166.51&167.69&$\displaystyle\matrix{167.97\cr 167.95}$&
  $\displaystyle\matrix{168.05\cr 168.03}$&167.44&169.05&169.56&169.80\\
\hline &&&&&&&&\\
  165.0&171.72&172.93&$\displaystyle\matrix{173.22\cr 173.20}$&
  $\displaystyle\matrix{173.30\cr 173.28}$&172.64&174.28&174.80&175.05\\
\hline &&&&&&&&\\
  170.0&176.92&178.17&$\displaystyle\matrix{178.47\cr 178.45}$&
  $\displaystyle\matrix{178.55\cr 178.53}$&177.84&179.52&180.05&180.30\\
\hline
\end{tabular}
\caption{\label{tab1}
Top quark mass relations for the $\overline{\rm MS}$, PS, $\overline{\rm PS}$,
and the pole mass at LO, NLO, NNLO, and ``large $\beta_0$'' accuracy in $\GeV$.
We fix the $\overline{\rm MS}$ mass to be $\overline{m}(\overline{m})=160\GeV$,
$165\GeV$, and $170\GeV$ and find the pole mass at LO, NLO, and NNLO from the
three-loop relation in Eq.~(\ref{msmass}). The PS and $\overline{\rm PS}$
masses are derived from the pole mass by using Eq.~(\ref{psmass}) (without or
with the $1/m$ contributions, respectively). We use the QCD coupling constant
$\alpha_s(m_Z)=0.119$,  $\mu=\overline{m}(\overline{m})$, and
$\mu_f=20\GeV$ for the factorization scale in the PS and $\overline{\rm PS}$
masses.}
\end{center}
\end{table}
}
In Fig.~\ref{fig8} we show the difference between the $\overline{\rm PS}$ and
the PS mass (in GeV) as a function of the factorization  scale $\mu_f$ (solid
line) at $\mu=15\GeV$. It is interesting to observe that the non-abelian part
of the difference between the $\overline{\rm PS}$ and the PS mass
(dotted line in Fig.~\ref{fig8}) can be as large as $200\MeV$. But the recoil
correction cancel in part the non-abelian one and therefore the final
difference is not more than $50\MeV$. The dependence of
$m_{\overline{\rm PS}}-m_{\rm PS}$ on the renormalization scale at
$\mu_f=30\GeV$ is given in Fig.~\ref{fig8} by the dashed curve.

\section{NNLO results}
\subsection{The scheme with the pole mass}
The top quark cross section at LO, NLO, and NNLO is shown in Fig.~\ref{fig9}
as a function of the center-of-mass energy. For the top quark pole mass we use
$m_t=175.05\GeV$, for the top quark width $\Gamma_t=1.43\GeV$, and for the QCD
coupling constant $\alpha_s(m_Z)=0.119$~\cite{PDG}. Different values
$\mu=15\GeV$, $30\GeV$, and $60\GeV$ for the renormalization scale are
selected because they roughly correspond to the typical spatial momenta for
the top quark. For solving the Schr\"odinger equation we use the program
written by one of the authors~\cite{Yakovlev:1999ke}. Note that we do not take
into account an initial photon radiation which would change the shape of the
cross section. This can be easily included in the Monte Carlo simulation.

The NNLO curve modifies the line shape by the amount of $20-30\%$ which is as
large as the NLO correction. It also shifts the position of the $1S$ peak by
approximately $600\MeV$. These large shifts of the peak position were expected.
As we discussed above (and is well-known in the
literature~\cite{Beneke:1994sw,Vainshtein}), the pole mass definition suffers
from the renormalon ambiguity. The top quark pole mass cannot be defined better
than $O(\Lambda_{\rm QCD})$. Large NNLO corrections and a large shift of the
$1S$ resonance can spoil the top quark mass measurement at the NLC.

\subsection{The scheme with the PS mass}
In this subsection we discuss the calculation scheme using the PS mass. We
redefine the pole mass through the PS mass using the relation given in
Eq.~(\ref{psmass}) without the $1/m$ contributions and then use the PS mass as
an input parameter for our numerical analysis at LO, NLO, and NNLO. In
Fig.~\ref{fig10} we show the top quark cross section expressed in terms of
$m_{\rm PS}(\mu_f)$ at LO, NLO, and NNLO (like in Fig.~\ref{fig9}) as a
function of the center-of-mass energy. We take
$m_{\rm PS}(\mu_f=20\GeV)=173.30\GeV$ which corresponds to the pole mass
$m=175.05\GeV$ according to Table~\ref{tab1}. In looking at Fig.~\ref{fig10}
we observe an improvement in the stability of the position of the first peak
in comparison to the previous analysis as we go from LO to NLO to NNLO.
Actually, for the three values $\mu=15\GeV$, $30\GeV$, and $60\GeV$ we obtain
the maxima of the $1S$ peak for NNLO at $s_{\rm max}=347.32\GeV$, $347.41\GeV$,
and $347.48\GeV$ while the maximal values are given by $R_{\rm max}=1.379$,
$1.184$, and $1.088$, respectively. This demonstrated that a large variation
in the renormalization scale $\mu$ gives rise only to a shift of about
$160\MeV$ for the $1S$ peak position at NNLO while the variation for
$R_{\rm max}$ is still large.

\subsection{The scheme with the $\overline{\rm PS}$ mass}
\begin{figure}
\centerline{\epsfig{file=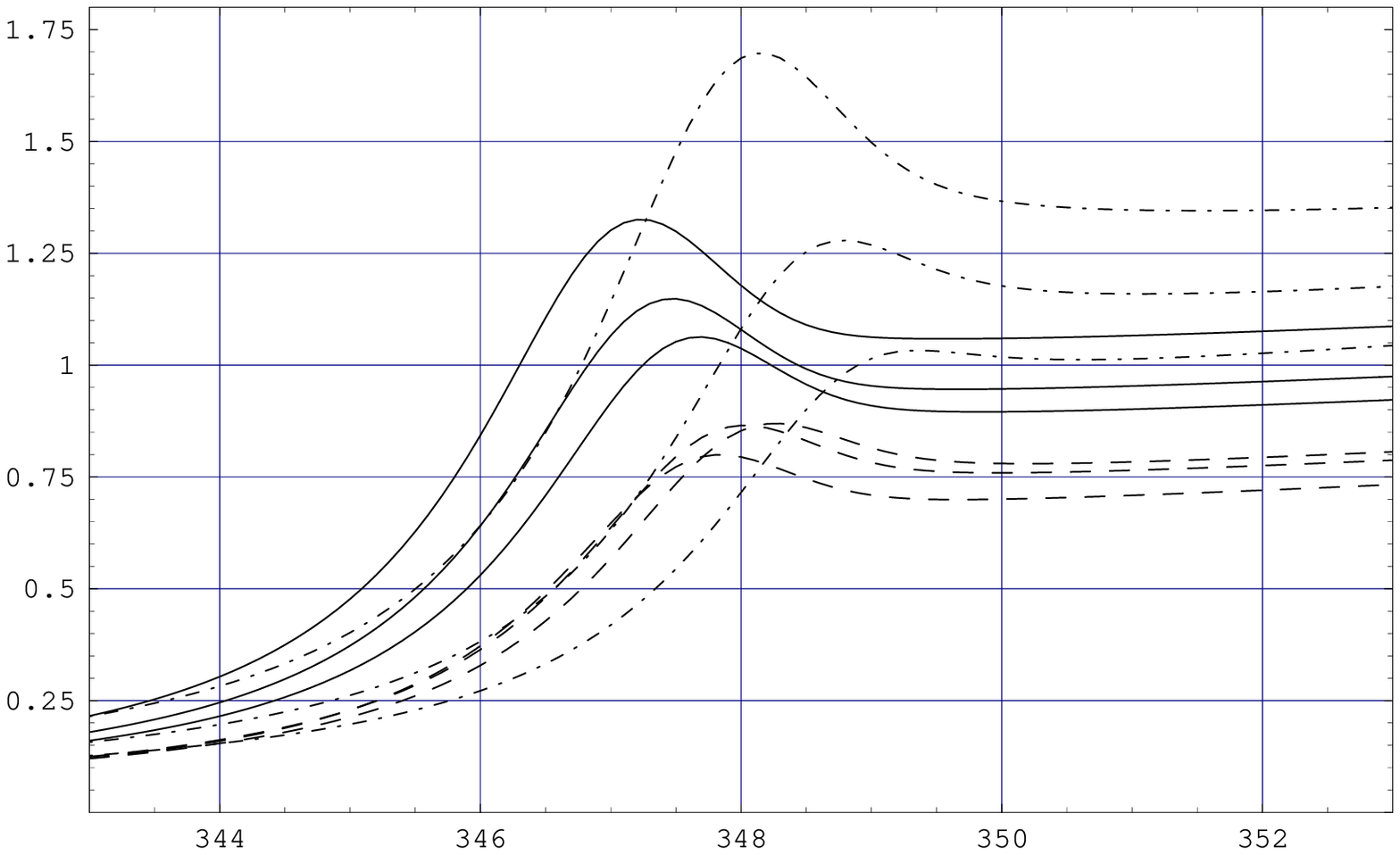,scale=0.7}}
\caption{\label{fig9}The scheme using the pole mass:
shown is the relative cross section $R(e^+e^-\to t\bar t)$ as a function of
the center-of-mass energy in $\GeV$ for the LO (dashed-dotted lines), NLO
(dashed lines), and NNLO (solid lines) approximation. We take the value
$m_t=175.05\GeV$ for the pole mass of the
top quark, $\Gamma_t=1.43\GeV$ for the top quark width,
$\alpha_s(m_Z)=0.119$ and different values
$\mu=15\GeV$, $30\GeV$, and $60\GeV$ for the renormalization scale.}
\end{figure}

\begin{figure}
\centerline{\epsfig{file=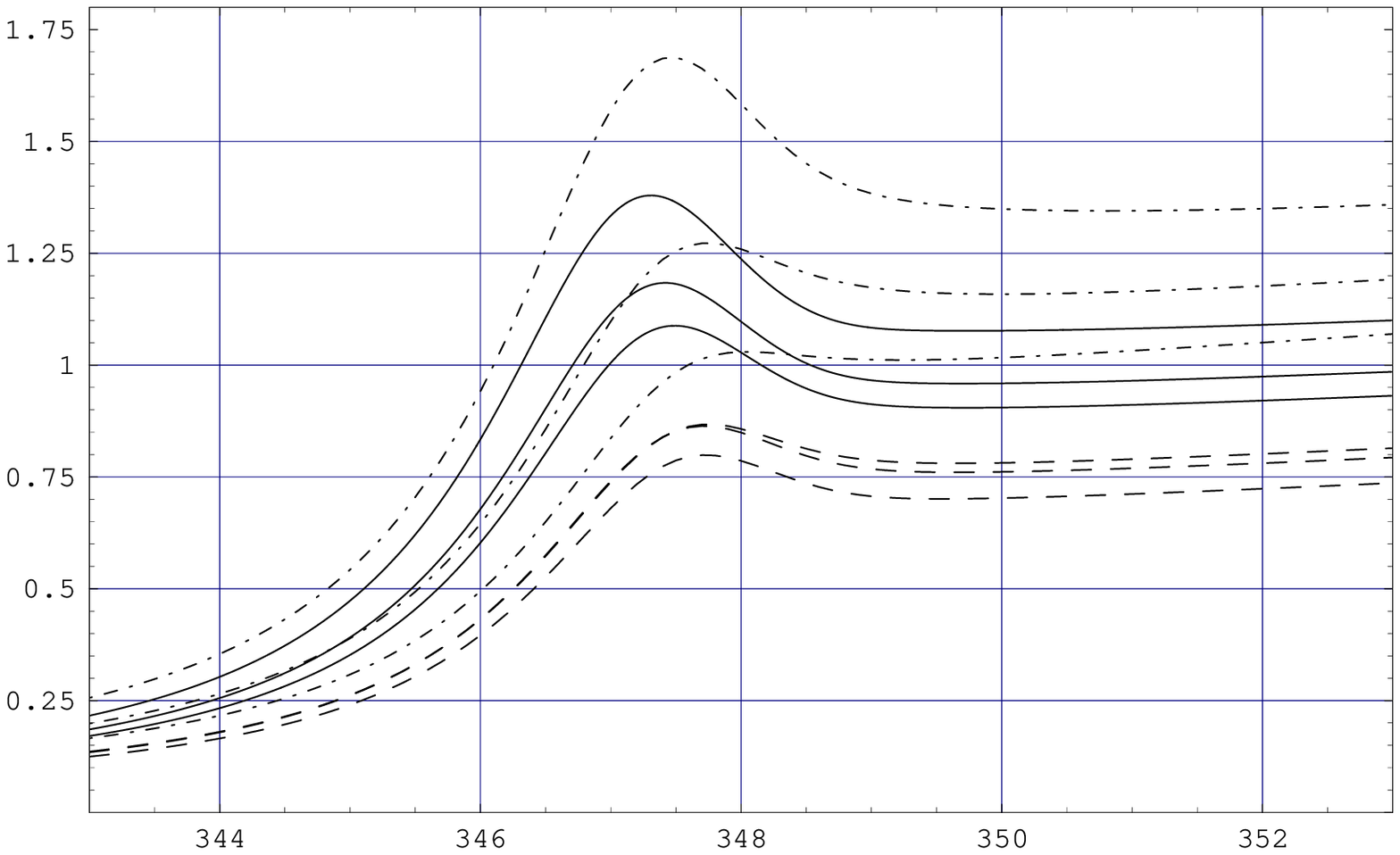,scale=0.7}}
\caption{\label{fig10}The scheme using the PS mass:
shown is the relative cross section $R(e^+e^-\to t\bar t)$ as a function of
the center-of-mass energy in $\GeV$ for the LO (dashed-dotted lines), NLO
(dashed lines), and NNLO (solid lines) approximation. We take the value
$m_{\rm PS}=173.30\GeV$ for the PS mass of the
top quark, $\Gamma_t=1.43\GeV$ for the top quark width,
$\alpha_s(m_Z)=0.119$ and different values
$\mu=15\GeV$, $30\GeV$, and $60\GeV$ for the renormalization scale.}
\end{figure}

\begin{figure}
\centerline{\epsfig{file=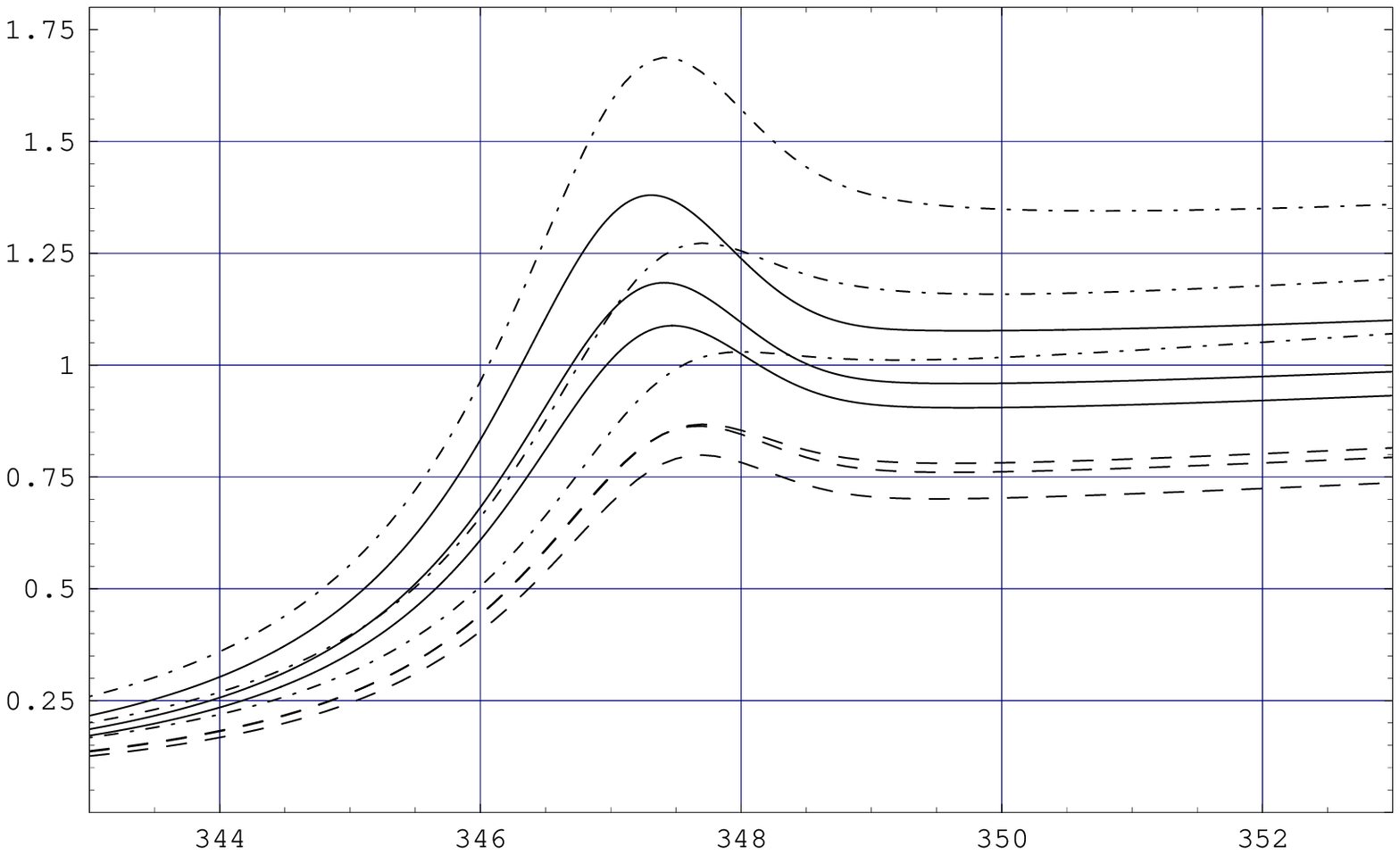,scale=0.7}}
\caption{\label{fig11}The scheme using the $\overline{\rm PS}$ mass:
shown is the relative cross section $R(e^+e^-\to t\bar t)$ as a function of
the center-of-mass energy in $\GeV$ for the LO (dashed-dotted lines), NLO
(dashed lines), and NNLO (solid lines) approximation. We take the value
$m_{\overline{\rm PS}}=173.28\GeV$ for the $\overline{\rm PS}$ mass of the
top quark, $\Gamma_t=1.43\GeV$ for the top quark width,
$\alpha_s(m_Z)=0.119$ and different values
$\mu=15\GeV$, $30\GeV$, and $60\GeV$ for the renormalization scale.}
\end{figure}

In this subsection we discuss the calculation scheme where we use the
$\overline{\rm PS}$ mass. We redefine the pole mass through the
$\overline{\rm PS}$ mass by using the relation given in Eq.~(\ref{psmass}) and
then use the $\overline{\rm PS}$ mass as an input parameter for the numerical
analysis at LO, NLO, and NNLO. In Fig.~\ref{fig11} we show the top quark cross
section expressed in terms of $m_{\overline{\rm PS}}(\mu_f)$ at LO, NLO, and
NNLO (like in Fig.~\ref{fig9}) as a function of the center-of-mass energy. We
take $m_{\overline{\rm PS}}(\mu_f=20\GeV)=173.28\GeV$. Again we observe a very
good stability of the position of the first peak as we go from LO to NLO to
NNLO, similar to the one observed for the PS mass case. Studying the size of
the NNLO corrections to the peak positions we conclude that the current
theoretical uncertainty of the determination of the PS mass from the $1S$ peak
position is about $100-200\MeV$.

\section{Conclusion and discussions}
We discussed the so-called potential subtracted (PS) mass suggested in
Ref.~\cite{Beneke} as a definition of the quark mass alternative to the pole
mass. In contrast to the pole mass, this mass is not sensitive to the 
non-perturbative QCD effects. We have derived recoil corrections to the
relation of the pole mass with the PS mass. The main result for this is
Eq.~(\ref{psmass}). In addition, we demonstrated that, if we use the PS or
the $\overline{\rm PS}$ mass in the calculations, the perturbative predictions
for the cross section become much more stable at higher orders of QCD (shifts
are below $0.1\GeV$). This understanding removes one of the obstacles for an
accurate top mass measurement and one can expect that the top quark mass will
be extracted from a threshold scan at NLC with an accuracy of about
$100-200\MeV$.\\[12pt]
{\bf Acknowledgements:}
We are grateful to R.~Akhoury, E.~Yao, M.~Beneke, A.~Hoang, D.~Gerdes, and
U.~Bauer for valuable discussions. O.Y.\ acknowledges support from the US
Department of Energy (DOE). S.G.\ acknowledges a grant given by the DFG, FRG, 
he also would like to thank the members of the theory group at the Newman Lab
for their hospitality during his stay.

\newpage


\end{document}